\documentclass[useAMS,usenatbib]{mn2e}
\usepackage{graphicx}
\usepackage{epsfig}
\usepackage{subfigure}
\usepackage{color}
\usepackage{xcolor}
\usepackage{multirow}
\usepackage{threeparttable}
\usepackage{natbib}
\usepackage{amsmath,upgreek,amssymb}

\title[Constraining the mass of galaxies in A611]{Separating Galaxies from the Cluster Dark Matter Halo in Abell~611}

\author[A. Monna et al.]
{\parbox{\textwidth}{ A. Monna$^{1,2}$\thanks{E-mail:
amonna@usm.uni-muenchen.de},
S. Seitz$^{1,2}$,
M. J. Geller$^{3}$,
A. Zitrin$^{4,5}$,
A. Mercurio$^{6}$,
S. H. Suyu$^{7,8}$
M. Postman$^{9}$,
D. G. Fabricant$^{3}$,
H. S. Hwang$^{10}$,
A. Koekemoer$^{9}$
}\vspace{0.4cm}\\
\parbox{\textwidth}{$^{1}$University Observatory Munich, Scheinerstrasse 1, 81679 Munich, Germany\\
$^{2}$Max Planck Institute for Extraterrestrial Physics, Giessenbachstrasse, 85748 Garching, Germany\\
$^{3}$Harvad-Smithsonian Astrophysical Observatory, 60 Garden St., Cambridge, MA 02138\\
$^{4}$Cahill Center for Astronomy and Astrophysics, California Institute of Technology, MS 249-17, Pasadena, CA 91125, USA\\
$^{5}$Hubble Fellow\\
$^{6}$INAF/Osservatorio Astronomico di Capodimonte, Via Moiariello 16, I-80131 Napoli, Italy\\
$^{7}$Max-Planck-Institut f{\"u}r Astrophysik, Karl-Schwarzschild-Str. 1, 85741 Garching, Germany\\
$^{8}$Institute of Astronomy and Astrophysics, Academia Sinica, P.O. Box 23-141, Taipei 10617, Taiwan\\
$^{9}$Space Telescope Science Institute, 3700 San Martin Drive, Baltimore, MD 21208, USA\\
$^{10}$School of Physics, Korea Institute for Advanced Study, 85 Hoegiro, Dongdaemun-gu, Seoul 02455, Korea\\
}}
\begin{document}
 \date{}


\maketitle

\label{firstpage}

\begin{abstract}
We investigate the mass content of galaxies in the core of the galaxy cluster Abell 611. We perform a strong lensing analysis of the cluster core and use velocity dispersion measurements for individual cluster members as additional constraints. Despite the small number of multiply-imaged systems and cluster members with central velocity dispersions available in the core of A611, the addition of velocity dispersion measurements leads to tighter constraints on the mass associated with the galaxy component, and as a result, on the mass associated with the dark matter halo. Without the spectroscopic velocity dispersions, we would overestimate the mass of the galaxy component by a factor of $\sim1.5$, or, equivalently, we would underestimate the mass of the cluster dark halo by $\sim5\%$. We perform an additional lensing analysis using  surface brightness (SB) reconstruction of the tangential giant arc. This approach improves by up to a factor $\sim10$ the constraints on the mass parameters of the 5 galaxies close to the arc.  
The resulting parameters are in good agreement with the $\rm \sigma-r_{tr}$ scaling relation derived in the pointlike analysis. The galaxy velocity dispersions resulting from the SB analysis are consistent at the $1\sigma$ confidence level with the spectroscopic measurements. In contrast the truncation radii for 2-3 galaxies depart significantly  from the galaxy scaling relation and suggest differences in the stripping history from  galaxy to galaxy.
\end{abstract}

\begin{keywords}
dark matter, galaxy cluster, galaxy haloes, gravitational lensing.
\end{keywords}

\section{Introduction}

Galaxies, and to a larger extent, clusters of galaxies, are dominated by  \textit{dark matter} (DM). Although DM cannot be observed directly, it can be detected through its gravitational effects. Thus gravitational lensing is a powerful tool for investigating the distribution of dark matter \citep[e.g. see][]{Schneider2003, Bartelmann2010,Kneib2011}. 
 Lensing allows a direct probe of the total projected mass density of the lens. In the case of strong lensing (SL) by galaxy clusters,  the location and redshift of sets of multiple images enable mapping of the mass-density distribution. To disentangle the cluster-scale dark halo component from the galaxies' contribution, additional constraints  sensitive to only one of the two components are necessary \citep{Eichner2013,Monna2015}.\\
\begin{figure*}
 \centering
 \includegraphics[width=17.5cm]{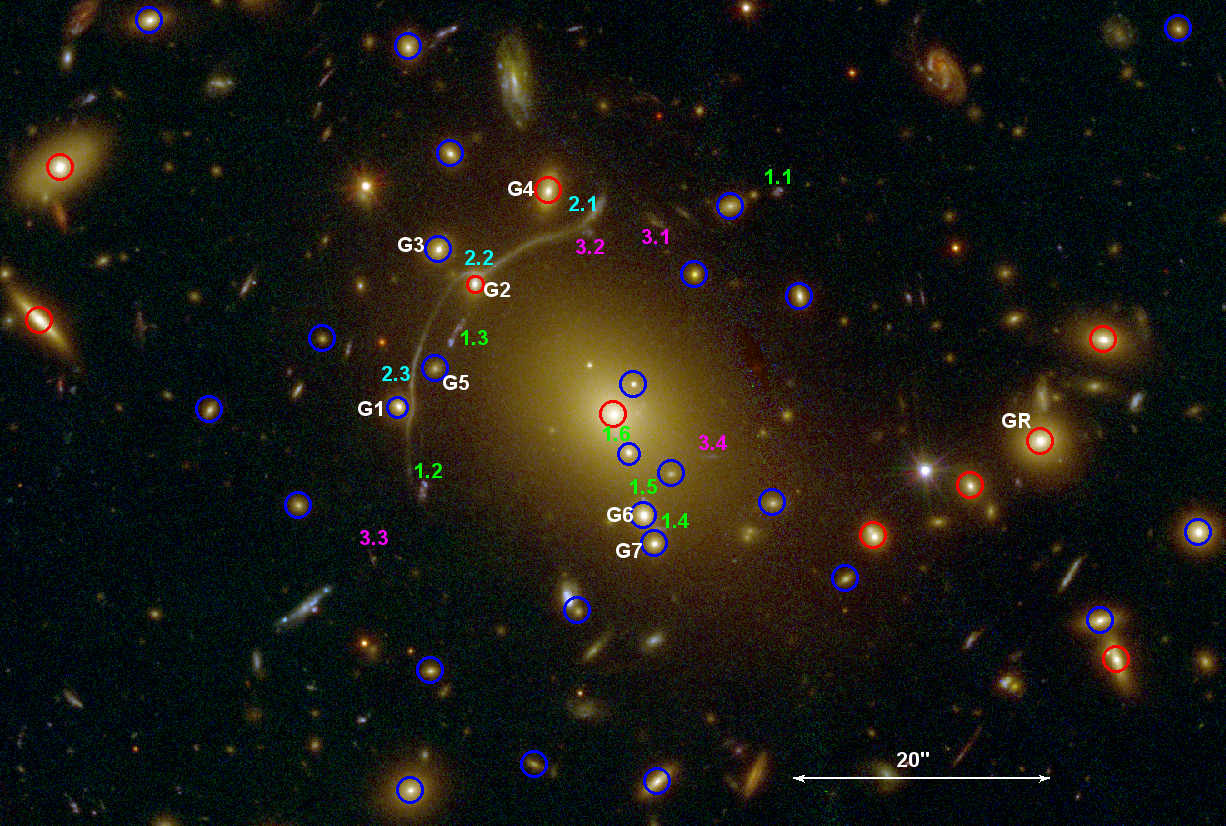}
 \caption{\small Colour composite image of the core ($\sim1.5\arcmin\times1.5\arcmin$) of A611 created using the CLASH HST dataset: Blue=F435W+F475W; Green=F606W+F775W+F814W+F850LP; Red=F105W+F110W+F140W+F160W. 
 Blue circles mark the cluster members included in the SL analysis. Red circles mark the galaxies with a measured central velocity dispersions.  ``GR'' indicates the galaxy used as reference for the luminosity scaling relation.  ``G1'' to ``G7'' are the galaxies we model individually. 
 The 3 multiply lensed systems used in the SL analysis are labelled in green (system 1), cyan (system 2) and magenta (system 3). North is up and East is left. }
         \label{fig:a3611rgb}
 \end{figure*}
 A common assumption is that the DM content of galaxies scales in proportion to their light \citep[e.g. see][]{Koopmans2006}. Luminosity scaling relations allow  estimates of some properties of the galaxy dark halos including their mass or central velocity dispersion and their extent. Weak and strong lensing analyses reveal that the extent of a galaxy's DM halo also depends on its  environment  \citep[e.g. see][]{Narayan1998, Geiger1999, Limousin2007, Halkola2007}. Galaxy-galaxy lensing analyses show that typical radii of dark matter halos are of the order of hundreds of kpc \citep[][]{Brimioulle2013,Limousin2007}. In denser environment, like galaxy clusters \citep{Limousin2007, Limousin2009}, galaxies are stripped during interactions with each other and with the smooth extended cluster dark matter halo.  Simulations predict that galaxies in the cluster core should be strongly truncated \citep{Merritt1983,Limousin2009} in  agreement with lensing results. However uncertainties in the measurements of the truncation radii are  large \citep{Nat2002,Limousin2007, Halkola2007, Donnarumma2011}.

In parametric strong lensing analysis the mass of a galaxy is often represented by a velocity dispersion.
For a singular isothermal sphere, $M_{tot}\propto\sigma^2 r_{tr}$ \citep{Elisa2007}, where $\sigma$ is the central velocity dispersion and $r_{tr}$ is the halo truncation radius. Galaxy velocity dispersions are inferred directly from their luminosity through the Faber-Jackson relation ($L_e\propto\sigma_0^\alpha$) \citep{Faber1976}. The respective mass is then estimated through the luminosity -- velocity dispersion -- mass scaling relations. However, large scatter in the Faber-Jackson relation inherently introduces modelling biases in lensing analyses.

Direct measurement of the velocity dispersions of individual cluster members allows a more direct estimate of their total mass, independent of the lensing signal. Spectroscopic velocity dispersion measurements are thus useful for separating the galaxy component from the global cluster DM halo. \citet{Monna2015} show that using velocity dispersion measurements in a SL analysis  sets stronger constraints on the galaxy dark matter halos and breaks the internal degeneracy between their mass profile parameters. By using Hectospec \citep{Fabricant2005, Fabricant2013} velocity dispersions for $\sim15$ cluster members in the core of the galaxy cluster Abell 383, \citet{Monna2015} improve  constraints on the galaxy luminosity scaling relations by 50\%. Furthermore,  surface brightness reconstruction of the giant arc in Abell 383 measures the extent of the DM halos of some cluster members near the arc.  \\
 \begin{table}
\caption{CLASH Photometric Dataset: column (1) filters, column (2) HST instrument,  column (3) $5\sigma$ magnitude depth within $0.6\arcsec$ aperture.}
\centering
\footnotesize
\begin{tabular}{|c|c|c|}
\hline
\hline
Filter&Instrument& 5$\sigma$ Depth\\
      \hline
f225w &   WFC3/UVIS & 25.4\\
f275w &   WFC3/UVIS & 25.6\\
f336w &   WFC3/UVIS & 26.0\\
f390w &   WFC3/UVIS & 26.5\\
f435w &  ACS/WFC    & 26.3    \\    
f475w &  ACS/WFC    & 26.7   \\  
f606w &  ACS/WFC    & 27.0     \\
f775w &  ACS/WFC    & 26.2     \\
f814w &  ACS/WFC    & 26.7  \\
f850lp&  ACS/WFC   &   25.9  \\
f105w &  WFC3/IR   &     26.9     \\
f110w &  WFC3/IR   &     26.9     \\
f125w &  WFC3/IR   &     26.8     \\
f140w &  WFC3/IR   &     26.9     \\
f160w &  WFC3/IR   &     26.7     \\
\hline                  
\end{tabular}
\label{tab:phot}
\end{table}

\begin{table*}
\centering
\caption{\small List of  cluster members with measured velocity dispersion in the core of A611. Col.1 gives the ID, Col. 2 and 3 give the Ra and Dec in degrees, Col. 4 the spectroscopic redshift and Col. 5 the \texttt{auto\_mag} extracted with Sextractor in the F814W filter. Col. 6 provides the effective radius measured with \texttt{GALFIT} in the HST/F814W band, and  Col.7 lists the measured central velocity dispersion. }
\footnotesize
\begin{tabular}{|c|c|c|c|c|c|c|}
\hline
\hline
ID & RaJ2000 & DecJ2000& z$_{sp}$& mag$_{\rm F814W}$ & R$_{\rm eff}$\,(kpc) &$\sigma_{sp}$\, (km/s)\\
\hline
  BCG    &	120.23676	&	36.05657	&0.287&	16.6	&40.0 &$	330	\pm	19$	\\
  GR&	120.22530	&	36.05598        &0.291&	19.2	&5.4& $	185	\pm	25$	\\
  G2  	&	120.24033	&	36.05950	&0.283&	20.0	&3.4& $	124	\pm	45$	\\
  G4  	&	120.23850	&	36.06141	&0.281&	19.9	&1.9& $	214	\pm	72$	\\
  243 	&	120.22977	&	36.05394	&0.287&	19.9	&1.3& $	251	\pm	37$	\\
  446 	&	120.21320	&	36.06816	&0.286&	20.5	&2.3& $	293	\pm	105$	\\
  123 	&	120.25097	&	36.04291	&0.283&	19.0	&9.0& $	251	\pm	24$	\\
  533 	&	120.25350	&	36.07586	&0.290&	19.4	&3.5& $	249	\pm	56$	\\
  489 	&	120.25670	&	36.06833	&0.284&	19.4	&2.9& $	235	\pm	42$	\\
  248 	&	120.22717	&	36.05502	&0.295&	20.7	&1.3& $	231	\pm	81$	\\
  159 	&	120.25952	&	36.04721	&0.283&	19.4	&2.4& $	228	\pm	41$	\\
  345 	&	120.25213	&	36.05861	&0.291&	19.1	&4.1& $	221	\pm	28$	\\
  316 	&	120.22361	&	36.05819	&0.297&	19.6	&2.2& $	217	\pm	43$	\\
  402 	&	120.25158	&	36.06192	&0.288&	19.1	&4.2& $	208	\pm	19$	\\
  380 	&	120.21555	&	36.06329	&0.287&	20.3	&2.0& $	178	\pm	44$	\\
  137 	&	120.23792	&	36.04691	&0.279&	20.4	&1.3& $	156	\pm	54$	\\
  512 	&	120.21346	&	36.06996	&0.278&	20.3	&2.2& $	139	\pm	54$	\\
\hline     
\hline     
  \end{tabular}
\label{tab:vel_disp}
\end{table*}


Here we investigate a second cluster, Abell 611 (A611 hereafter) at $z_{cl}\sim0.288$. Several previous SL analyses  derive a mass distribution for A611 focusing on different aspects of mass modelling through strong lensing. For example, \citet{Richard2010}   analyse 20 galaxy clusters from the LoCuSS survey, including A611, to constrain the total mass in cluster cores and to compare them with X-ray measurements.  \citet{Donnarumma2011}   combine SL and X-ray analyses of A611 and tested different approaches for modelling the mass associated with the cluster galaxy component. Within the cluster core (r$<100$ kpc), SL and X-ray mass measurements agree well, but in the outer region there are discrepancies.
The disagreement at large radius suggests an incorrect estimate of the relative
contributions of the baryonic and DM components to the cluster mass as a result of the degeneracy between the galaxy and the smooth large scale dark matter components. \citet{Newman2009,Newman2013a} combine kinematic data with strong and weak lensing   to model the mass distribution of the cluster from the very central region out to well beyond the virial radius. They use measurement of the BCG velocity dispersion to constrain its stellar mass and thus to disentangle the baryonic and DM components in the very core. An NFW profile fails to describe the cluster DM alone; shallower profiles fit  the data better. More recently, \citet{Zitrin2015} presented a statistical weak and strong lensing analysis of the complete CLASH cluster sample, including A611. They examine the effect of different mass parametrizations on the resulting mass model and show that systematic differences of $40\%$ in the projected surface mass density can result from various lensing degeneracies.

We use a set of  spectroscopic velocity dispersions measured with Hectospec  \citep[see][]{Fabricant2005, Fabricant2013} mounted on the Multiple Mirror Telescope (MMT) to improve  constraints on the mass distribution of A611 and its mass components. We carry out a SL analysis for A611, both with and without the velocity dispersion measurements and examine  the impact of these measurements in constraining  both the mass in galaxies and in the dark matter halo. We also use the surface brightness morphology of the giant arc to refine  constraints on five cluster galaxies neighbouring the arc. 

\S \ref{sec:dataset} describes the photometric and spectroscopic datasets. 
\S \ref{sec:lensing}  describes the SL analysis, the mass components included in the mass model and the lensed systems used as constraints. The SL analysis results are in \S \ref{sec:A611}. We discuss the improvements resulting from the use of velocity dispersion measurements in the lensing analysis. \S \ref{sec:sb} discusses the surface brightness reconstruction of the giant tangential arc. We conclude in  \S \ref{sec:conclusions}. We assume a cosmological model with Hubble constant $H_0 = 70$ km s$^{-1}$ Mpc$^{-1}$ and density parameters  $\Omega_{\rm m} = 0.3$ and $\Omega_{\Lambda}=0.7$. Magnitudes are in the AB system.

\section{Photometric and Spectroscopic Dataset} 
\label{sec:dataset}
A611 was observed in 2012 during the Hubble Space Telescope (HST) Cycle 19, as part of the CLASH survey. It was observed with the HST Advanced Camera for Surveys (HST/ACS) and the Wide Field Camera 3 (HST/WFC3) UVIS and IR cameras providing deep photometry in 15  different HST filters, to a depth of roughly $\sim27$AB (3$\sigma$). The photometric dataset\footnote{available at http://archive.stsci.edu/prepds/clash/} was processed with the \texttt{Mosaicdrizzle} pipeline \citep[see][]{Koekemoer2011} to generate mosaic drizzled images with a common scale of 65mas/pixel, centred on the cluster. The field of view (FOV) is $\sim 3.5'\times3.5'$ in the ACS filters and $\sim2'\times2'$ in the WFC3IR images.   In Tab.~\ref{tab:phot}  we list the filters of the photometric data with the respective $5\sigma$ depths.  We generate multi-band photometric catalogues of  fluxes extracted within $0.6\arcsec$ (9 pixels) diameter aperture using \texttt{SExtractor} 2.5.0 \citep{Bertin1996} in dual image mode. We use the weighted sum of all of the WFC3IR images as the detection image.

We measured redshifts and velocity dispersions with the Hectospec fiber spectrograph mounted on the MMT
\citep{Fabricant2005}. Hectospec has 300 fibers deployable over a 1 degree field. The instrument has 1.5$^{\prime\prime}$ fibers and the spectra cover the wavelength range 3500 - 9150\,\AA. The resolution is of 5.5\,\AA\, FWHM, which corresponds to 105 km/s at 6000\,\AA.  We acquired data under variable conditions with typical seeing of 0.9\arcsec on
February 9, April 5, and April 9-10, 2010, on November 22-23, 2011, and on
October 8 and November 28, 2013 \citep{Fabricant2013}. The typical integration time was 1 hour.
All redshifts are published in \citet{Lemze2013}.

Within 1.5$^{\prime}$ of the cluster centre, the Hectospec data provide redshifts for
27 cluster members and central velocity dispersions for seventeen of these members. We include a central velocity dispersion here only when its uncertainty is $\lesssim 40$\% of the dispersion.

We extract  the velocity dispersions from the Hectospec spectra by applying
the University of Lyon Spectroscopic analysis Software7
\citep[UlySS;][]{Koleva2009}. Single age stellar population
models calculated with the PEGASE-HR
code \citep{Leborgne2004} and the MILES stellar
library \citep{Sanchez2006} provide the basis for fitting
the observed spectrum. We limit the spectral fitting range to $4100-5500$ Angstrom rest-frame wavelengths where we obtain the smallest velocity dispersion errors and the most stable velocity dispersions \citep{Fabricant2013}.
Based on the measured
line spread function, we convolve models  to the wavelength
dependent spectral resolution of the Hectospec
data \citep{Fabricant2013}. We convolve models that are parametrized
by age and metallicity with the
line-of-sight velocity dispersion; we then use $\chi^2$ minimization to determine the best-fit age, metallicity
and velocity dispersion.
\citet{Fabricant2013} discuss the details of these measurements. \citet{Zahid2015} 
carry out a further demonstration of the excellent agreement between these measurements and those derived for the same galaxies by the SDSS.

Following \cite{Jorgensen1995}, we correct the measured velocity dispersion $\sigma_{obs}$ observed with the Hectospec $1.5^{\prime\prime}$ aperture fibers, to estimate the galaxy central stellar velocity dispersion $\sigma_{sp}$ within the effective radius of the galaxy , $\rm R_{eff}$, according to
\begin{equation}
 \sigma_{\rm sp}=\sigma_{\rm obs}\left(\frac{\rm R_{eff}}{8\times d/2}\right)^{-0.04}\,,
\label{eq:sigma}
 \end{equation}
where  $d$ is the fiber aperture. 
We estimate the effective radii of cluster members  with \texttt{GALFIT} \citep{Peng2010} by fitting the surface brightness distribution of the galaxies in the HST/F814W filter with de Vaucouleurs profiles. Table\,\ref{tab:vel_disp} lists
all of the galaxies with measured velocity dispersions used in the  lensing analysis.

\section{Strong Lensing modeling}
\label{sec:lensing}

We model the mass distribution in the core  of A611  using  the  
software {\sc GLEE} developed by A. Halkola and S. Suyu \citep{Suyu2010,Suyu2012}. 
We use the observed positions of the multiple images as constraints throughout the analyses; thus we refer to these models as {\it pointlike}. Spectroscopic redshifts of the lensed sources, if available, are additional constraints.
The best fitting model is found through a simulated annealing minimization in the image plane. 
The most probable parameters and uncertainties for the cluster mass model are then obtained from  Monte Carlo Markov Chain (MCMC) sampling.

\subsection{Multiple images}
To reconstruct the mass distribution in the core of A611, we use three robust systems of multiple images as constraints \citep[Fig.\,\ref{fig:a3611rgb}; see][]{Newman2013a,Zitrin2015}. \\
{\bf System 1} is a quintuply lensed source spectroscopically confirmed at $z_{sp}=1.56$ (the redshift has been recently revised, see \citealt{Newman2013a}). A central image, embedded in the BCG light, is associated with this system and is included as a constraint, for a total of 6 multiple images.\\
{\bf System 2} is a giant tangential arc at a distance of $\sim18\arcsec$ from the BCG, passing in between 5 bright galaxies (G1-G5). We use the positions of the three brightest knots of the arc as constraints in the analysis. This system has a spectroscopic redshift of $z_{sp}=0.9$.\\
Finally {\bf system 3} is a quadruply lensed source, for which no spectroscopic redshift is available. Thus its redshift is a free parameter and is optimized around the photometric redshift $z_{phot}=1.54$ of the brightest lensed image (3.1) using a flat prior in the range [0.5, 2.5]. \\
The position of all the multiple images are listed in Tab.\,\ref{tab:multiple images} together with their respective redshifts.\\ 
We adopt errors of $1\arcsec$ on the position of the observed multiple images to account for uncertainties due to density fluctuations along the line of sight \citep[see ][]{Host2012,D'Aloisio}.
\begin{table}
\centering
\caption{\small Multiply imaged systems used to constrain the SL model of A611.
  The columns are: Col.1 is the ID; Col.2-3 Ra and Dec; Col.4  is the source redshift $\rm z_s$ which is spectroscopic for System 1 and 2, and photometric for system 3 ( \citealt{Zitrin2015}); Col. 5 is the final source redshift from the strong lensing model; col. 6 provide the difference between the observed and predicted position of each multiple image resulting from our final best cluster model (see Sec.\,\ref{sec:A611}). 
  }
\footnotesize
\begin{tabular}{|l|c|c|c|c|c|}
\hline
\hline
 Id & Ra & Dec & $\rm z_s$ & $\rm z_{sl}$& $\delta\theta (\arcsec)$\\
 \hline 
1.1   & 120.232260&   36.061430&   1.56 &- &1.1     \\ 
1.2   & 120.241820&   36.055075&   1.57 &- &0.6     \\ 
1.3   & 120.241110&   36.058144&   -    &- &0.4     \\ 
1.4   & 120.235610&   36.054100&   -    &- &0.1     \\ 
1.5   & 120.235950&   36.054732&   -    &- &0.2     \\ 
1.6   & 120.236680&   36.056140&   -    &- &0.8     \\ 
\hline                                                
2.1   & 120.237240&   36.060997&   0.91 &- & 1.0    \\ 
2.2   & 120.240480&   36.059643&   -    &- & 0.6    \\ 
2.3   & 120.242150&   36.057169&   0.86 &- & 0.6    \\ 
\hline                                                
3.1   & 120.235610&   36.060708&   1.54 &$1.68\pm0.20$ & 0.7    \\ 
3.2   & 120.237380&   36.060528&   1.12 &\arcsec & 0.3    \\ 
3.3   & 120.243160&   36.053450&   1.52 &\arcsec & 0.4    \\ 
3.4   & 120.234070&   36.055653&   -    &\arcsec & 1.1    \\ 
\hline     
\hline     
  \end{tabular}\\
\label{tab:multiple images}
\end{table}

\subsection{Cluster mass component}

{\bf Cluster smooth large scale halo}\\
We describe the smooth dark halo (DH) mass component of the cluster with a Pseudo Isothermal Elliptical Mass Distribution (PIEMD) profile \citep{Kassiola1993}.
Its projected  surface density is
 \begin{equation}
 \Sigma(R)= \frac{\sigma^2}{2 {\rm G}}\left(\frac{1}{\sqrt{r_{\rm c}^2+R^2}}\right),
 \end{equation}
where $\sigma$ and $\rm r_{\rm c}$ are the halo velocity dispersion and core radius. 
$R$ is the 2D radius, defined as $R^2=x^2/(1+e)^2+y^2/(1-e)^2$ for a profile with ellipticity $e=(1-b/a)/(1+b/a)$, where $b/a$ is the axis ratio. 
The asymptotic ($b/a\rightarrow1, r_{\rm c}\rightarrow0$) Einstein radius $\theta_E$ for this profile is
\begin{equation}
\theta_E=4\pi\left(\frac{\sigma}{c}\right)^2\frac{D_{ds}}{D_s}=\Theta_E\frac{D_{ds}}{D_s}
\end{equation}
where $\sigma$ is the halo velocity dispersion, $c$ is the speed of light, and $D_s$ and $D_{ds}$ are the distances to the lensed source and between the lens and the source, respectively. The  
 \textit{Einstein parameter} $\Theta_E$  is the Einstein radius for $D_{ds}/D_s=1$. 
We use  $\Theta_E$ as a parameter to describe the mass amplitude of the lens halo.
All the DH parameters are optimized using flat priors. The halo is initially centred on the BCG, and its position is optimized within 3 arcsec. The axis ratio  and  position angle (PA) vary  within [0,1] and $180^\circ$ respectively. The core radius varies within [0,60]\,kpc. The Einstein parameter $\Theta_E$ is optimized in the range $[4.5^{\prime\prime},65^{\prime\prime}]$
corresponding to a velocity dispersion of $\sim[400,1500]$\,km/s.\\

{\bf BCG}\\
\citet{Newman2009,Newman2013a,Newman2013b} perform a detailed analysis of the stellar mass profile of the BCG in A611 by combining SL and kinematic analyses. They model the BCG stellar mass profile using a dual pseudo elliptical isothermal profile \citep[dPIE, see][]{Elisa2007} to fit the surface brightness profile of the BCG.
The effective radius of the BCG is consistent with the truncation radius of the dPIE profile ($\rm r_{\rm tr,BCG}=46.2\pm3.4$ kpc  \citealt{Newman2013a,Newman2013b}). By combining lensing and kinematic analysis they estimate the amplitude of the dPIE profile  describing the BCG stellar mass. This amplitude is characterized by a central velocity dispersion  $\rm \sigma_{0,BCG}^*=164\pm33$\,km/s. \\
In our lensing model, we also describe the BCG stellar mass component by using a dPIE profile. For this mass component we adopt the parameters estimated by \citet{Newman2013a,Newman2013b} and we optimize them within their uncertainties using gaussian priors. 
\\

\begin{figure}
 \centering
 \includegraphics[width=8.5cm]{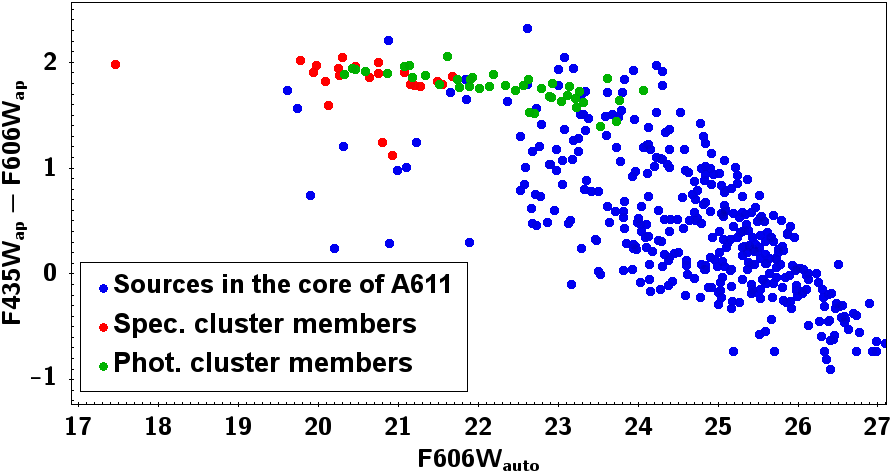}
 \caption{\small Colour-magnitude diagram for A611. We show all of the sources extracted in the core of A611 (blue), the cluster members with measured spectroscopic redshift (red), and  the photometric candidate cluster members (green) included in the SL model.}
         \label{fig:mag_col_diagr}
 \end{figure}

{\bf Cluster members}\\
Our analysis includes cluster members within a FOV of $\sim 1.5'\times1.5'$ centred on the cluster core (RAJ2000=$120.23676$	 and DECJ2000=$36.05657$).
Within this FOV there are  27 spectroscopically confirmed cluster members from the Hectospec survey with $\rm |z_{\mathrm{sp}}-z_{\mathrm{cl}}|<0.02$. 
Further candidate cluster members are selected photometrically. We select bright galaxies ($\rm m_{auto, F606w}\leq25$) on the cluster red sequence with $1.3\leq\rm m_{F435W}-m_{F606W}\leq2.3$ (Fig.\,\ref{fig:mag_col_diagr}).
We also require that these galaxies have  photometric redshift  $\rm |z_{ph}-z_{cl}|\leq0.03$. Photometric redshifts are estimated using the CLASH dataset (see Tab.\,\ref{tab:phot}),  based on the spectral energy distribution (SED) fitting code \texttt{LePhare}\footnote{http://www.cfht.hawaii.edu/~arnouts/lephare.html} \citep{Arnouts1999,Ilbert2006},
using  the COSMOS galaxy spectra \citep{Ilbert2009} as templates.

We select a total of 69 cluster members; 27 are spectroscopic members and 42 are photometric candidates.\\
Galaxies are  modelled with dPIE profiles \citep{Elisa2007}. This profile is characterized by a central velocity dispersion $\sigma$, a core radius $r_{\rm c}$ and a truncation radius $r_{\rm tr}$, marking the region where the density slope changes from $\rho\propto r^{-2}$ to $\rho\propto r^{-4}$.\\
 The projected surface mass density is
 \begin{equation}
 \Sigma(R)= \frac{\sigma^2}{2 {\rm G} R}\frac{r_{\rm tr}^2}{(r_{\rm tr}^2-r_{\rm c}^2)} \left(\frac{1}{\sqrt{1+\frac{r_{\rm c}^2}{R^2}}}-\frac{1}{\sqrt{1+ \frac{r_{\rm tr}^2}{R^2}}}\right)
 \label{eq:bbs}
 \end{equation}
 where $R^2=x^2/(1+e)^2+y^2/(1-e)^2$, as for the PIEMD mass profile.
The total mass is 
\begin{equation}
  M_{\rm tot}=\frac{\pi\sigma^2}{G}\frac{r_{\rm tr}^2}{r_{\rm tr}+r_{\rm c}}\,\,,
  \label{eq:m_bbs1}
\end{equation}
which, for $r_{\rm c}\rightarrow0$, reduces to
\begin{equation}
 M_{\rm tot}=\frac{\pi\sigma^2r_{\rm tr}}{G}\,\,.
 \label{eq:m_bbs}
\end{equation}
For a vanishing core radius, $r_{\rm tr}$ corresponds to the radius containing half of the total mass of the galaxy \citep[ see Appendix A3 in ][]{Elisa2007}.   
We adopt vanishing core radii for the cluster members, unless stated otherwise. Thus there are 2 free parameters associated with each galaxy,  $\sigma$ and $r_{\rm tr}$.

To reduce the total number of free parameters for the cluster members, we adopt luminosity scaling relations to derive the galaxy parameters $\sigma$ and $r_{\rm tr}$ \citep[e.g. see][]{Jullo2007, Eichner2013}. 
The Faber-Jackson relation \citep{Faber1976} allows estimation of the central velocity dispersion of elliptical galaxies directly from the observed photometry:
\begin{equation}
 \rm \sigma_i=\sigma_{GR}\left(\frac{L_i}{L_{GR}}\right)^{\delta}\,,
 \label{eq:FJ}
\end{equation}
where $\rm \sigma_{GR}$ and $\rm L_{GR}$ are the central velocity dispersion and luminosity of a fiducial reference galaxy. \\
The Fundamental Plane \citep[e.g. see][]{Dressler1987,Faber1987, Bender1992} also provides a luminosity scaling relation for the size of the halos. Following \citet{Hoekstra2003, Halkola2007, Limousin2007}  the truncation radii of galaxy halos scale with their luminosities according to
\begin{equation}
 \rm r_{tr,i}=r_{tr,GR}\left(\frac{L_i}{L_{GR}}\right)^{\alpha}\propto r_{tr,GR}\left(\frac{\sigma_i}{\sigma_{GR}}\right)^\frac{\alpha}{\delta}\,,
\label{eq:FP}
 \end{equation}
where  $\rm r_{tr,GR}$ and $\rm L_{GR}$ are the truncation radius and luminosity of the reference galaxy.\\
The total mass-to-light ratio for the cluster members is
\begin{equation}
 \rm \frac{M_{tot}}{L}\propto L^{\epsilon}\,,
\end{equation}
which, given Eq.\,\ref{eq:FP}, implies
\begin{equation}
 \rm M_{tot}\propto\sigma^{\frac{1}{\delta}(\epsilon+1)}\,.
\label{eq:mtot}
 \end{equation}
Combining Eq.\,\ref{eq:m_bbs}, \ref{eq:FP} and \ref{eq:mtot} we derive a relation among the exponents of the luminosity scaling relations
\begin{equation}
 2\delta+\alpha-\epsilon=1\,.
 \label{eq:exp}
\end{equation}

Measurements of the exponent $\delta$ from photometric analysis give value between 0.25 and 0.3 \citep[see e.g., ][]{Kormendy2013, Focardi2012}: measurements through strong and weak lensing analysis provide a value of $\delta=0.3$ \citep[see][]{Rusin2003, Brimioulle2013}. 
From our sample of galaxies with measured velocity dispersion in A611,  $\delta=0.3$ in the F814W band. Thus we adopt this value. Following our previous analysis, we assume a constant mass-to-light ratio ($\epsilon=0$) \citep[see][]{Eichner2013,Monna2015}, which then implies $\alpha=0.4$ from Eq.\,\ref{eq:exp}.  
The exponents of the scaling relations  are fixed to these values throughout our analysis. The constraints available in the field of the cluster are insufficient  for further investigation of the choice of values of the exponents. Throughout our analysis we do investigate the amplitude of the luminosity scaling relation by tuning the parameters $\sigma_{GR}$ and $\rm r_{tr,GR}$ of the reference galaxy. 
Thus the velocity dispersion $\sigma_i$ and  truncation radius $r_{i,tr}$ of the $i-$th cluster member follow the  scaling relation
\begin{equation}
 \rm \sigma_i=\sigma_{GR}\left(\frac{L_i}{L_{GR}}\right)^{0.3}\,\,r_{tr,i}=r_{tr,GR}\left(\frac{L_i}{L_{GR}}\right)^{0.4}
\end{equation}
where  $\sigma_{\rm GR}$ and $r_{\rm GR,tr}$ are the values for the reference galaxy GR.\\ 

As fiducial galaxy GR we adopt a bright elliptical ($b/a\sim0.9$, $\rm f814w_{\rm best}=19.2$)  $\sim30^{\prime\prime}$ west of the BCG, slightly outside the critical lines for a source at redshift $\rm z_s=2$. This galaxy has a measured velocity dispersion $\rm \sigma_{sp,GR}=185\pm25$\,km/s (Table\,\ref{tab:vel_disp}). We optimize the velocity dispersion through the lensing analysis with a gaussian prior consistent with the measurement uncertainties.
The GR truncation radius is a free parameter optimized with a flat prior in the range $[0.3\arcsec, 20\arcsec]$. \\
The cluster members with measured $\rm \sigma_{sp}$ are assigned their central velocity dispersions fixed to the spectroscopic measurement throughout the modelling. Thus we assume that the measured central stellar velocity dispersion $\rm \sigma_{sp}$ is a robust estimate of  the central velocity dispersion of the galaxy halo. \\
In addition seven galaxies located very close to the tangential arc and to the central multiple images of system 1 (G1 to G7,  see Fig.\,\ref{fig:a3611rgb}), are individually optimized. Two of these galaxies, G2 and G4, have measured $\rm \sigma_{sp}$ that are optimized with a gaussian prior within the measurement errors. 
The central velocity dispersions for the other galaxies (G1, G3, G5, G6 and G7) are optimized with a flat prior in the range $[10,400]$\,km/s. For these  5  cluster members, the truncation radii are free parameters optimized within $[0.3\arcsec,20\arcsec]$. \\
The position, axis ratio $b/a$, and PA  of all the galaxies are fixed to the values extracted from the photometry in the F814W filter. Only for the galaxies G1 to G5, the axis ratio and PA are optimized with a gaussian prior.   

Finally, we also allow for an external shear component to  account for the  large scale environmental contribution to the lensing potential. 

\section{Pointlike model}
\label{sec:A611}
\begin{table}
\centering
\caption{\small Final parameters of the mass components of A611 resulting from modelling the mass distribution with (``w/$\sigma$'') or without (``wo/$\sigma$'') the measured velocity dispersions. The uncertainties are 68\% confidence limits from the MCMC sampling.
x and y position are given in kpc with respect to the BCG. Core and truncation radii are given  in kpc, velocity dispersions are in km/s and Einstein radii parameters are in arcseconds. PA are in radians measured counter-clockwise from the west direction.}
\footnotesize
\begin{tabular}{|l|c|c|c|c|c|}
\hline
\hline
  & wo/$\sigma$& w/$\sigma$ \\
  \hline             															
 DH & &\\         															
 \hline            															
 x              &   $	0.0	\pm3.5$	&	 $	-2.2	\pm2.6$		\\
 y              &   $	5.6	\pm3.5	$	&	 $	4.3	\pm2.6	$		\\
 b/a            &   $	0.8	\pm0.1	$	&	 $	0.8	\pm0.1$		\\
PA              &   $	2.1	\pm0.2$	&	 $	2.2	\pm0.1	$		\\
 $\theta_e$     &   $	18	\pm2$	&	 $	20	\pm1$		\\
 $\rm r_{core}$ &   $	20	_{-	9	}^{+	13	}$	&	 $	25	_{-	7	}^{+	9	}$		\\
 \hline            															
BCG &&\\         															
 \hline            															
$\sigma$      &   $	167\pm31$	&	 $	167\pm31$		\\
$\rm r_{core}$  &   $	1.3	$	&	 $	1.3$		\\
$\rm r_{\rm tr}$    &   $	44\pm8$	&	 $	43\pm9$		\\
 \hline            															
 GR &&\\         															
 \hline            															
 $\sigma $    &   $	250\pm63$	&	 $	186\pm27$		\\
 $\rm r_{\rm tr}$   &   $	55	_{-	29	}^{+	22	}$	&	 $	42	_{-	26	}^{+	28	}$		\\
 \hline            															
 G1 &&\\         															
 \hline            															
b/a             &   $	0.8	\pm0.2$	&	 $	0.8\pm0.2$		\\
 PA             &   $	1.3\pm0.2	$	&	 $	1.3	\pm0.2$		\\
 $\sigma$     &   $	270\pm 71$	&	 $	212\pm49$		\\
  $\rm r_{\rm tr}$  &   $	25	_{-	19	}^{+	38	}$	&	 $	42	_{-	27	}^{+	32	}$		\\
  \hline            															
 G2 &&\\         															
 \hline            															
b/a             &   $	0.7	\pm0.2$	&	 $	0.7	\pm0.2	$		\\
 $\sigma $    &   $ 234\pm59$	&	 $	132\pm40$		\\
   $\rm r_{\rm tr}$ &   $	33	_{-	24	}^{+	34	}$	&	 $	41	_{-	28	}^{+	30	}$		\\
 \hline            															
 G3 &&\\         															
 \hline            															
 b/a            &   $	0.8\pm0.2	$	&	 $	0.8	\pm0.2$		\\
 $\sigma$     &   $	257\pm81$	&	 $	264\pm53$		\\
   $\rm r_{\rm tr}$ &   $	37	_{-	26	}^{+	34	}$	&	 $	39	_{-	28	}^{+	32	}$		\\
 \hline            															
 G4 &&\\         															
 \hline            															
b/a             &  $	0.5	\pm0.2	$	&	 $	0.5	\pm0.2	$		\\
 $\sigma$     &  $212\pm74$	&	 $195\pm36$		\\
   $\rm r_{\rm tr}$ &  $	41	_{-	29	}^{+	31	}$	&	 $	34	_{-	25	}^{+	35	}$		\\
 \hline            															
 G5 &&\\         															
 \hline            															
$\sigma$     &   $	186\pm65$	&	 $	83\pm21$		\\
  $\rm r_{\rm tr}$  &   $	35	_{-	26	}^{+	35	}$	&	 $	40	_{-	27	}^{+	31	}$		\\
 \hline            															
 G6 &&\\         															
 \hline            															
b/a             &   $	0.8	\pm0.2	$	&	 $	0.8	\pm0.2$		\\
$ \sigma$     &   $	186\pm47$	&	 $	186\pm28$		\\
  $\rm r_{\rm tr}$  &   $	41	\pm31$	&	 $	40	_{-	30	}^{+	32	}$		\\
 \hline            															
 G7 &&\\         															
 \hline            															
b/a             &   $	0.9\pm0.1	$	&	 $	0.9	\pm0.1$		\\
PA              &   $	0.8\pm0.2$	&	 $	0.8\pm0.2$		\\
$\sigma $    &   $195\pm62$	&	 $	156\pm33$		\\
  $\rm r_{\rm tr}$  &   $	37	_{-	27	}^{+	33	}$	&	 $	42	\pm30$		\\
\hline
$\gamma$ &$0.01_{-0.01}^{+0.02}$&$0.005\pm0.003$\\
  \hline     
\hline     
  \end{tabular}\\
\label{tab:wo_wsigma_comparisons}
\end{table}

\begin{figure}
 \centering
 \includegraphics[width=8.5cm]{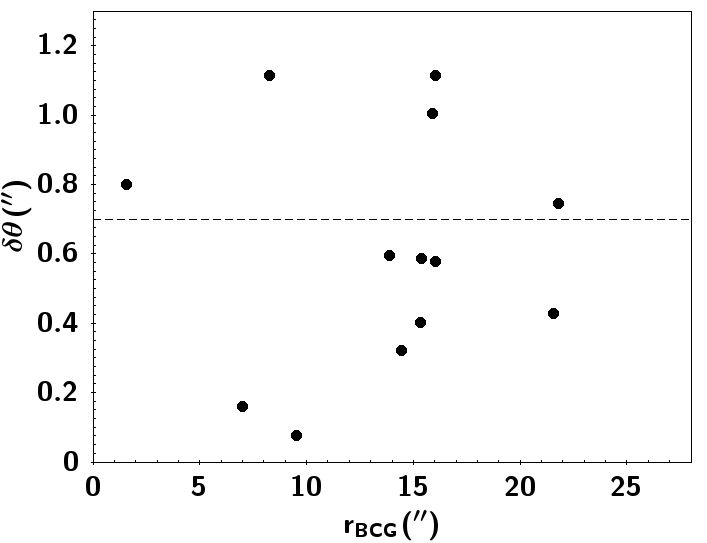}
 \caption{\small Distance $\delta\theta$ between the predicted and observed multiple images as a function of the distance of the respective multiple images from the BCG. The multiple images are reproduced with mean accuracy of 0.7\arcsec.}
         \label{fig:rms}
 \end{figure}

Given the model in Section 3, we perform a $\chi^2$ minimization analysis on the position of the observed multiple images. Then we run MCMC chains to obtain the final best model and the relative uncertainties. The final best model has $\chi^2=0.7$ and reproduces the multiple images with a mean accuracy of $0.7^{\prime\prime}$. Fig.\,\ref{fig:rms}  shows the distance $\delta\theta$ between the predicted and the observed positions of the multiple images as a function of the projected distance of the lensed image from the BCG. These $\delta\theta$ are in Tab.\,\ref{tab:multiple images}, together with the model prediction of the source redshift for system 3. System 3 has a mean rms of 0.6\arcsec  for the multiple image reproduction and a redshift $\rm z_{sl}=1.7\pm0.2$, consistent with the photometric redshift.\\
The model yields a cluster dark halo  centred at $x=-0.5\pm0.6$ arcsec, $y=1.0\pm0.6$ arcsec with respect to the BCG, its axis ratio is $b/a=0.8\pm0.1$ with P.A.=$126^{\circ}\pm6^{\circ}$.
The 
core radius is $25\pm8$ kpc and the Einstein parameter $\Theta_E=20.2\pm1.4$ arcsec corresponding to $\rm\sigma= 837\pm29$\,km/s.
For a source at $z_{\rm s}=2$, the halo has a fiducial Einstein radius $\theta_E=15.7\pm1.1$ arcsec.
The total mass of the cluster within the Einstein radius $\theta_E$ is $M_{\rm tot}=3.8\pm0.1\times10^{13}M_\odot$. The external shear component has magnitude of $\gamma=0.5\pm0.3\times10^{-2}$.

The final velocity dispersion for the reference galaxy is $\rm \sigma_{GR}=186\pm28$\,km/s and its truncation radius is  $r_{\rm tr,GR}=42\pm26$ kpc. 
The galaxy scaling relation is then
\begin{equation}
\sigma=\rm 186\pm28\,km/s\left(\frac{r_{tr}}{42\pm26\,kpc}\right)^{3/4}\,.
\label{eq:ls}
\end{equation}
Although the error in $r_{\rm tr}$ is large, the relation implies that galaxies are tidally truncated in the cluster core consistent with previous  analyses \citep[e.g.][]{Nat2002, Limousin2007, Donnarumma2011, Monna2015}. 
 These results are in excellent agreement with \citet{Donnarumma2011}, who perform a detailed strong lensing and X-ray analysis of A611. They test different approaches to modelling the mass associated with the cluster galaxy component. 
They individually optimize the mass parameters of galaxies which have strong impact on the observed lensing features just as we do in our analysis. In {\rm case 6} of their analysis,  the velocity dispersion and truncation radius of six galaxies close to the lensed systems are individually optimized in the ranges [90, 190]\, km/s and [2, 35] kpc, respectively for $\sigma$ and $\rm r_{tr}$. In contrast with our approach, their scaling relation reference galaxy GR has a fixed truncation radius of 43 kpc (in agreement with results from galaxy-galaxy lensing presented in \citet{Nat2009}) and has $\rm \sigma_{tr,GR}$ optimized in the range [120, 200]\,km/s. 
They perform the lensing analysis in the source plane and obtain a total reduced $\chi^2_{src}=0.7$. The final velocity dispersion for their reference galaxy GR is $\sigma_{GR}=150\pm18$\,km/s,  consistent at the $1\sigma$ level with our results. We also find overall agreement  for the galaxies that we individually optimized. However, in both analyses the mass parameters have large uncertainties (see Fig.\,\ref{fig:sb} in the next section).  \\
Table \ref{tab:wo_wsigma_comparisons}  summarizes the parameters for all of the cluster mass components, including the individually optimized galaxies, G1 to G7.  \\
\begin{table}
\centering
\caption{\small 2D projected mass  for the DH and galaxy mass components of A611.   We extract the mass  enclosed within the Einstein parameter ($\Theta_E\sim20^{\prime\prime}$) and within a larger radius of $50^{\prime\prime}$, covering the cluster core.
 Masses are in $\rm 10^{13}M_\odot$. }
\footnotesize
\begin{tabular}{|l|c|c|}
\hline
\hline
  & wo/$\sigma$& w/$\sigma$ \\
 \hline             															
 DH & &\\         															
 $\rm M_{DH}(<20^{\prime\prime})$ &$2.7\pm0.1$&$2.9\pm0.1$\\
 $\rm M_{DH}(<50^{\prime\prime})$ &$6.6\pm0.7$&$7.0\pm0.6$\\
 \hline       
  Galaxies & &\\         															
 $\rm M_{gal}(<20^{\prime\prime})$ &$0.5\pm0.2$&$0.3\pm0.1$\\
 $\rm M_{gal}(<50^{\prime\prime})$ &$1.8\pm0.4$&$1.2\pm0.5$\\
 \hline  
 \end{tabular}
\label{tab:masses}
\end{table}

\begin{figure}
 \centering
 \includegraphics[width=8.5cm]{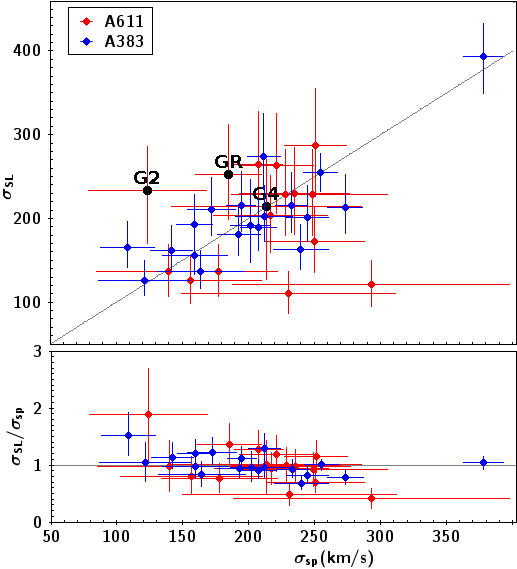}
 \caption{\small Spectroscopic velocity dispersions ($\rm \sigma_{sp}$) versus the prediction from lensing. 
We show the relation for A611 (red) along with the individually optimized objects (black).
For comparison we also show the results for  A383 (in blue) from  \citet{Monna2015}. Note the tighter relation for A383.}
         \label{fig:vel}
 \end{figure}

\begin{figure*}
 \centering
 \includegraphics[width=18cm]{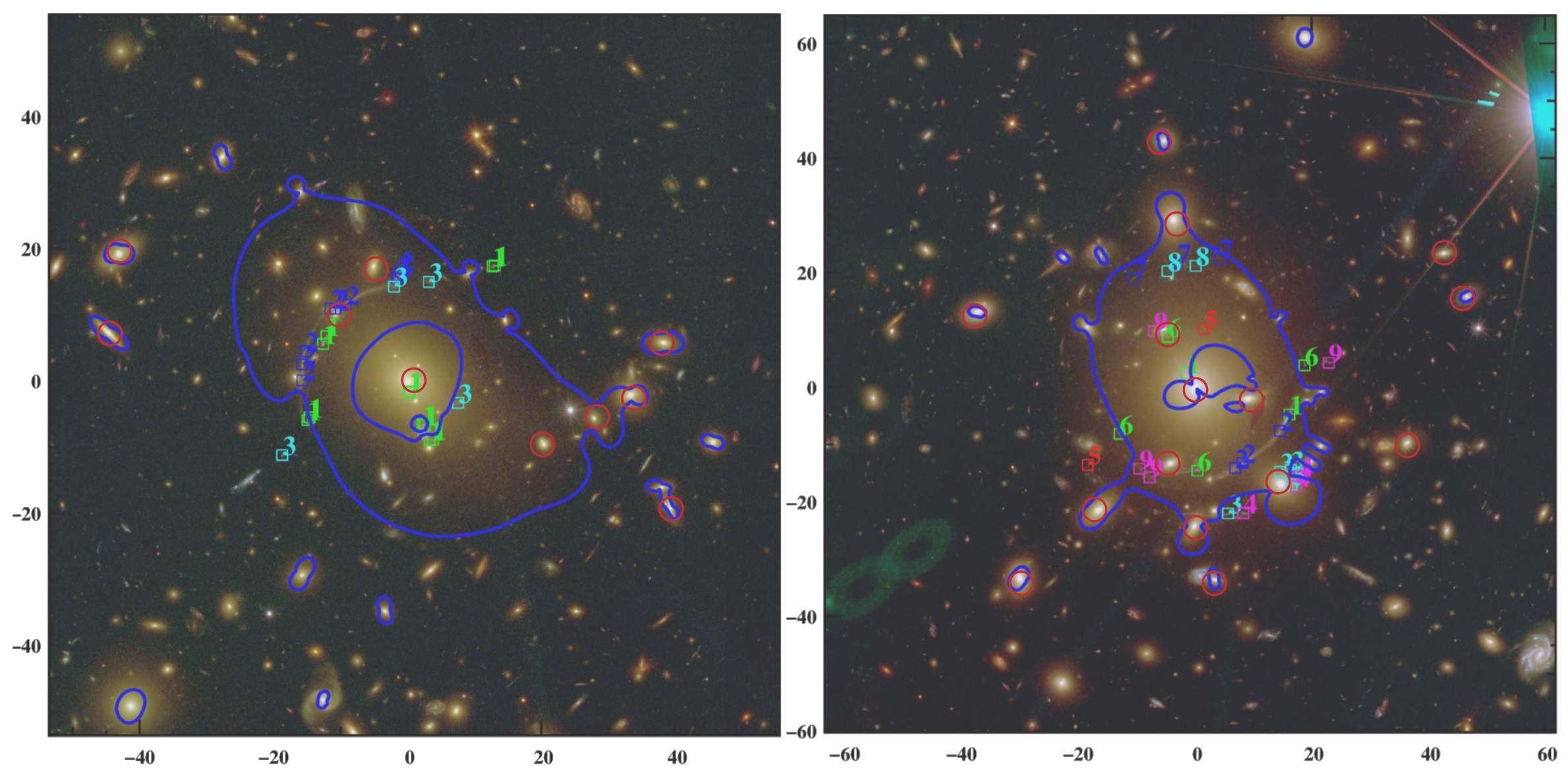}
 \caption{\small HST colour composite images of A611  (z=0.288; left) and A383  (z=0.189; right). We show the multiple image (numbered), the cluster members with measured $\rm \sigma_{sp}$ (red circles), and the critical lines for a source at redshift $\rm z_s=2$ (blue). Axis labels are in arcseconds with respect to the BCG.}
         \label{fig:gal_fov}
 \end{figure*}

To examine the impact of incorporating the measured velocity dispersions of the 17 cluster members in the lensing analysis, we model the cluster core without using the $\rm \sigma_{sp}$  (``wo/$\sigma$''). 
The best model  wo/$\sigma$  predicts the multiple images positions with a mean accuracy of 0.8\arcsec; with the measured velocity dispersions we obtain  $\rm rms_{img}=0.7$\arcsec.
Constraints on the parameters of the mass components are similar in both models (Tab.\,\ref{tab:wo_wsigma_comparisons}). However the velocity dispersions $\rm \sigma_{sl}$ predicted by lensing show some deviations from  the available spectroscopic measurements $\rm \sigma_{sp}$. 
Fig.\,\ref{fig:vel} shows the measured $\sigma_{sp}$ versus the value predicted from lensing ($\rm \sigma_{sl}$) for the model $wo/\sigma$. For comparison we also plot the results for A383 in \citealt{Monna2015}.  In A383  the spectroscopically measured $\rm \sigma_{sp}$ and the lensing prediction agree well.  The difference between A611 and A383 probably depends on the sample of cluster members for which we have $\rm \sigma_{sp}$.   In the case of Abell\,383 \citep{Monna2015}, there are 8 cluster members with $\rm \sigma_{sp}$ inside the cluster  critical lines, and another 5 slightly outside. Thus in A383 we constrain the galaxy truncation radii with an error of $\sim50\%$. In A611   we derive much weaker constraints on the truncation radii because the sample of cluster members with measured $\rm \sigma_{sp}$ includes  fewer galaxies (only 5 inside the critical lines) which have a strong impact on the lensing potential.
Furthermore A611 has the fewer robust SL features identified in the core of the cluster than A383 does. Fig.\,\ref{fig:gal_fov} shows the two clusters to highlight the difference between them.  The figure  marks the galaxies with measured $\rm \sigma_{sp}$ over a field of $50"\times50"$ centred on the BCG along with the multiple images and the critical curves. \\
The use of velocity dispersion measurements of cluster members  properly weights the galaxy contribution to the total mass of the cluster.  
The lensing analysis $wo/\sigma$ predicts $\rm r_{\rm tr}=55^{+29}_{-22}$\, kpc and $\rm \sigma_{GR,wo/\sigma}=250\pm63$\,km/s for GR. This velocity dispersion is overestimated by a factor of 1.4 with respect to the spectroscopic measurement $\rm \sigma_{sp,GR}=185\pm25$\,km/s. The total mass associated with GR is $\rm M_{tot,GR}^{wo/\sigma}=2.5\times10^{12}M_{\odot}$  compared with $\rm M_{tot,GR}^{w/\sigma}=1.1\times10^{12}M_{\odot}$ estimated using the measured $\rm \sigma_{sp}$ values in the analysis.\\ 
Overestimation of  the mass of GR  translates into a global overestimate of the total galaxy mass component in the cluster.  Table\,\ref{tab:masses} lists the projected mass for  the cluster galaxy contribution and for the large scale dark halo. The mass associated with the cluster members is over-estimated by a factor of $\sim1.5$ without  use of the velocity dispersions in the analysis. Consequently the mass associated with the large scale dark halo is underestimated by $\sim5\%$ in the model ``wo/$\sigma$''. 
In spite of the small sample of cluster members with measured velocity dispersions in A611, these spectroscopic measurements allow proper weighting of the mass associated with the galaxies relative to the large scale dark matter halo.
 
 \section{Surface Brightness Reconstruction of the giant arc}
 \label{sec:sb}
 In the pointlike analysis, we individually optimized some of the cluster members which reside very close to multiple images and thus probably contribute significantly to the local lensing distortion. However, the mass profiles for these galaxies are actually poorly constrained by the pointlike modelling.\\

\begin{figure*}
 \centering
 \includegraphics[width=17cm]{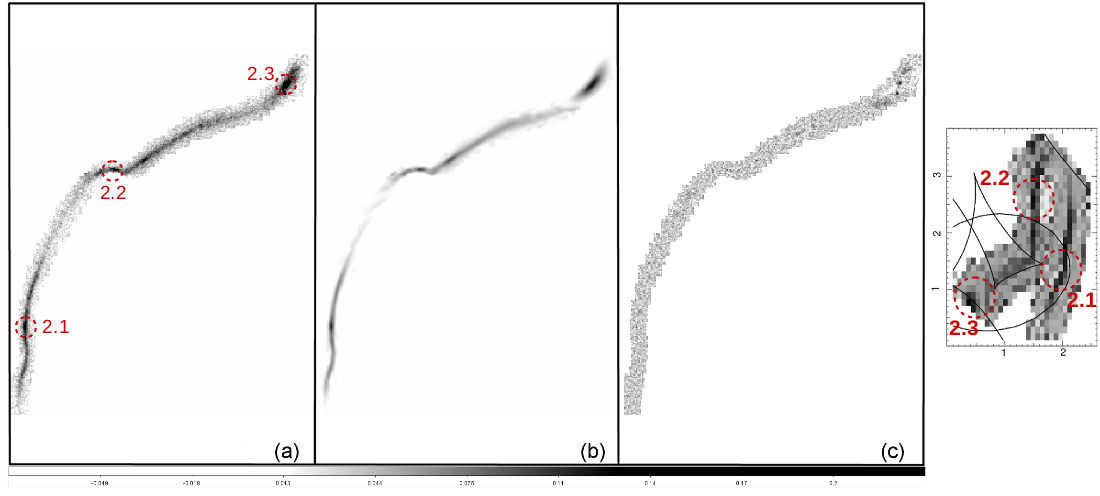}
 \caption{\small Surface brightness reconstruction of the giant arc in the core of A611. Panel (a) is  a cutout ($\sim16"\times20"$) of the arc in the HST/ACS/F775W filter; Panel (b) is the arc reconstruction resulting from the SB lensing modelling; Panel (c) shows their residuals. On the right side we show the reconstruction of the source, which is at $z_{sp}=0.91$. The source cut out has size of $\sim4\arcsec\times2.5\arcsec$, the scale shown is in arcseconds. The black lines are the caustic for the source redshift. The red dotted circles mark substructures identified in the arc (Panel (a)) and their respective position in the source plane. See text for more details.  
}
         \label{fig:sb}
 \end{figure*}

Here we use the surface brightness (SB) reconstruction of the giant arc 
 to improve the  constraints on the mass parameters of the galaxies, G1 to G5, that are close to the arc by using the entire pixel to pixel lensing information encoded in the arc.
 The analysis is performed in the HST/ACS/F775W image. We subtract the neighbouring cluster members from the image of the arc using \texttt{GALFIT} \citep{Peng2010} thus removing  light that would otherwise contaminate the arc SB reconstruction. 
 We start the SB modelling from the final best model obtained in the pointlike analysis. 
 The SB reconstruction of giant arcs is a powerful tool for placing strong constraints on the mass distribution of the lens, but it is only effective in the local
neighbourhood of the reconstructed arc. The pixel to pixel constraints associated with the giant arc are of the order of $10^3$ higher than the pointlike constraints associated with the other multiple image systems used in the analysis. Thus the minimization and the MCMC analyses are dominated by the constraints from the arc. Throughout the SB analysis we must fix the parameters of the mass components which do not play a major role in the local distortion of the giant arc. In Appendix \ref{sec:A1} we show the MCMC resulting from the pointlike analysis. The parameters describing the mass component of galaxies G1-G5 have no degeneraies with the parameters of the large scale dark halo, the BCG and the reference galaxy GR. Thus through the SB reconstruction of the giant arc, the
parameters describing the DH, BCG and GR are fixed to the values of the final best model of the pointlike analysis. Only the five galaxies G1-G5 close to the giant arc  (see Fig.\,\ref{fig:a3611rgb}), are individually optimized in the SB reconstruction. Their ellipticity, position angle, velocity dispersions and truncation radii are free parameters.  Given the larger number of constraints provided by the SB analysis,   we also release the core radii for these galaxies  to infer a more detailed profile of the  local mass distribution. Two of these galaxies, G2 and G4, have measured velocity dispersions (see Tab.\ref{tab:vel_disp}) optimized within the spectroscopic error with a gaussian prior.
We perform the SB reconstruction through a linear inversion method \citep{Warren2003} which reconstructs the pixellated brightness distribution of the source, with regularization of its intensity, through a Bayesian analysis \citep{Suyu2006}.\\  
 The first model resulting from the SB reconstruction has a large reduced $\rm \chi^2_{SB}=1.6$. For consistency with the pointlike analysis, where we adopt larger uncertainties in the multiple images position of 1\arcsec, at this stage of the SB analysis we increase the pixel noise associated with the HST/ACS/F775W image by a factor of $\sqrt{1.6}$. \\
Fig.\ref{fig:sb}  shows the results of the surface brightness reconstruction of the arc in the F775W filter as well as the reconstructed source. The giant arc is  reconstructed well(panel b in Fig.\,\ref{fig:sb}) with $\rm \chi^2_{SB}=0.8$ on the pixel intensities and with residuals lower then $10^{-3}$ (panel c in Fig.\,\ref{fig:sb}). Large residuals remain close to the upper multiple image 2.3, where a bright compact object is clearly identified. Our reconstruction does not reproduce this object as part of the lensed system. This compact object is unlikely to be part of the lensed source. Otherwise it would be possible to identify such a substructure close to the other multiply lensed structures of the arc as well (i.e. close to image 2.1 and 2.2). On the right side of Fig.\,\ref{fig:sb} we show the source reconstruction of the lensed system. We obtain a good SB reconstruction of the giant arc in the image plane. In the source plane, however, the substructures of the arc have an offset of $\sim1.5-2.5$ arcseconds, which corresponds to $\sim 10$ kpc at the redshift of the source.\\
  Tab.\,\ref{tab:sb}  summarizes the resulting mass parameters for  galaxies G1-G5 optimized through this analysis with their respective $1\sigma$ uncertainties.
 The halo axis ratio b/a and position angles are consistent with the results from the pointlike analysis, as well as with the values measured from the photometry. Only G4 obtains a considerably larger axis ratio b/a=1 with respect to the b/a=0.5 measured from the photometry and resulting from the pointlike model.  
The velocity dispersions for G2 and G4 are both consistent with the  spectroscopic measurements (see Tab.\ref{tab:vel_disp}), although G4 has  an higher value  compared with the estimates from the pointlike analysis. The other galaxies obtain velocity dispersions which depart significantly from the previous pointlike results.\\

\begin{figure*}
  \centering
  \includegraphics[width=12cm]{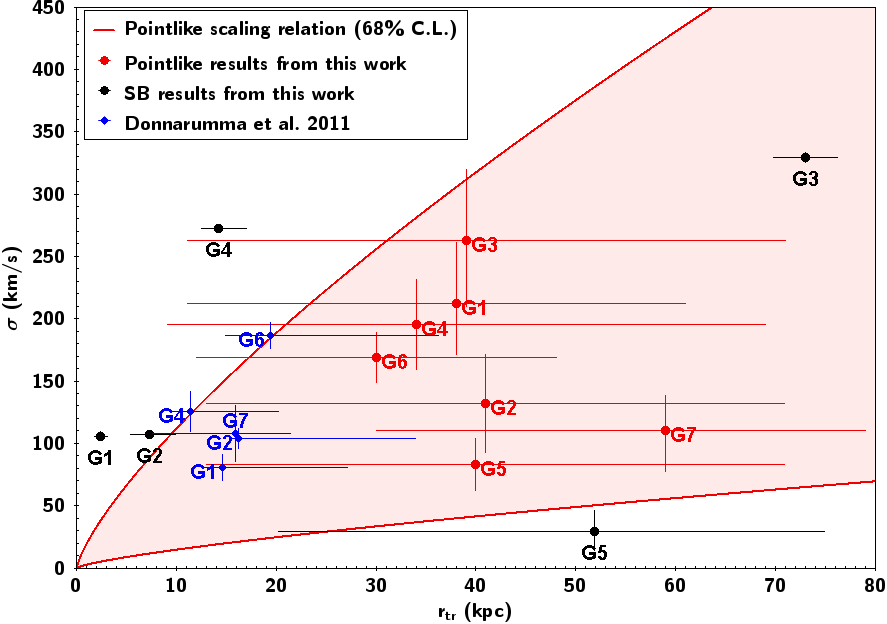}
  \caption{\small Velocity dispersions and truncation radii  for galaxies G1 to G5 resulting from the SB (black) and the pointlike (red) modelling. The red shaded region shows the cluster galaxy scaling relation derived from the pointlike analysis. In blue we plot results for the same galaxies from the pointlike SL analysis performed by \citet{Donnarumma2011}. The galaxies G2, G3 and G5 are consistent at the $1\sigma$ confidence level: galaxies G1 and G4 deviate by 2$\sigma$ or more.}
          \label{fig:scaling_law_sb}
  \end{figure*}

Overall the SB modelling yields improved constraints on the galaxy truncation radii and masses.
The truncation radii for G3, G4 and G5 agree at the $1\sigma$ level with the pointlike estimate. For  G1 and G2  we obtain quite small radii ($<15 $kpc) indicating that the
dark matter halos of these objects are highly truncated. The total mass associated with each galaxy is consistent within the 1$\sigma$ uncertainties with the mass estimated through the pointlike analysis, but now the masses are better constrained by a mean factor of $70\%$.
Fig.\,\ref{fig:scaling_law_sb} compares $\sigma$ and $\rm r_{tr}$ derived through the surface brightness reconstruction   with the pointlike  results.
At the  $1\sigma$ confidence level, the galaxies  are consistent with the scaling relation derived in the pointlike analysis, except for G1 and G2 which deviate substantially from the relation.  Fig.\,\ref{fig:scaling_law_sb} also shows results from \citet{Donnarumma2011} for comparison.
The \citet{Donnarumma2011} results  for galaxies G1-G5 are generally consistent with our scaling relation and with the results from the pointlike analysis. However,  they predict smaller  $\sigma$ and $\rm r_{tr}$ for these galaxies as a result of the smaller range adopted for the parameter optimization (  [120, 200] km/s for velocity dispersions and  [2, 35] kpc  for truncation radii). 
Nevertheless  overall both approaches predict that these galaxies are highly truncated. 
  According to the SB reconstruction, G1 has $\rm r_{\rm tr,G4}=2\pm1$\,kpc,  only twice its effective radius ($\rm R_{eff,G1}=1 kpc$) as measured in the HST/F814W image. This galaxy has a total mass $\rm M_{tot,G1}=2.0\pm0.5\times10^{11} M_{\odot}$.
Comparing  its total mass with a prediction for field galaxies \citep[e.g.][]{Brimioulle2013}, G1 has probably been stripped of $99\%$ of its original dark matter halo. 
G2 and G4 have also apparently lost most of their dark halos through stripping processes ($98\%$ and $97\%$ respectively); G3 and G5  have lost $85\%$ and $90\%$, respectively. 
Variations in the stripped mass may be explained by differences in the stripping processes resulting from  different orbits through the cluster \citep{Warnick2008}. 
\begin{table}
\centering
\caption{\small Final parameters of the mass components describing galaxies G1-G5 for the pointlike model and the SB model. The uncertainties are the 68\% confidence levels from the MCMC sampling. The radii are in kpc, the velocity dispersions are in km/s, the PA are in degrees measured counter-clockwisefrom the west direction. The masses are in $\rm 10^{12} M_{\odot}$.}
\footnotesize
\begin{tabular}{|l|c|c|c|c|c|}
\hline
\hline
  & Pointlike model& SB model \\
 \hline 
 G1 &&\\         															
 \hline            															
b/a             &   $	0.8	\pm0.2$	&$ 0.8	\pm0.1$		         \\
 PA             &   $	75	\pm11$	&$ 71	\pm4$		         \\
 $\sigma$     &   $	212.	_{-41}	^{+49}$	 & $  105	\pm3$   	 \\
  $\rm r_{\rm tr}$  &   $	38.	_{-27}	^{+32}$	&  $ 2\pm1$	 \\
$\rm r_{c}$  & $-$&$0.1\pm0.1$\\
$\rm M_{tot}$  & $1.2_{-1.2}^{+1.6}$&$ 0.20\pm0.05$\\
  \hline            															
 G2 &&\\         															
 \hline            															
b/a             &  	 $	0.7	\pm	0.2	$&$ 0.8\pm	0.1$		\\
 PA             &   $	[86]$	&$ 93	\pm6$		         \\
 $\sigma $    &  	 $132	\pm39.$	     &$107\pm2$          	\\
   $\rm r_{\rm tr}$ &  	 $41	_{-28}	^{+30}$	&$7\pm3$	\\
$\rm r_{c}$  & $-$&$0.7\pm0.2$\\
$\rm M_{tot}$  &  $0.5_{-0.5}^{+0.7}$& $0.6\pm0.2$\\
\hline            															
 G3 &&\\         															
 \hline            															
 b/a            &  	 $	0.8	\pm0.2	$&$ 0.6\pm0.1$		\\
 PA             &   $	[45]$	&$ 41	\pm1$		         \\
 $\sigma$     &  	 $263	_{-53}^{+57}$	&$329\pm4$	\\
   $\rm r_{\rm tr}$ &  	 $39	_{-28}	^{+32}$	&$73\pm3 $	\\
$\rm r_{c}$  & $-$&$0.1\pm0.1$\\
$\rm M_{tot}$  &  $2.0_{-2.0}^{+2.4}$& $5.8\pm0.3$\\
\hline            															
 G4 &&\\         															
 \hline            															
b/a             & 	 $	0.5	\pm0.2$		&$1^{+0.0}_{-0.01}$\\
 PA             &   $	[82]$	&$ 83	\pm3$		         \\
 $\sigma$     & 	 $195	\pm36$		  &$279\pm1$      \\
   $\rm r_{\rm tr}$ & 	 $34	_{-25}	^{+35}$		&$14\pm3$\\
$\rm r_{c}$  & $-$&$4\pm1$\\
$\rm M_{tot}$  &  $1.0_{-1.}^{+1.3}$& $0.8\pm0.1$\\
\hline            															
 G5 &&\\         															
 \hline            															
$\sigma $     &   	 $83\pm21$&$30\pm15$		        \\
  $\rm r_{\rm tr}$  &   	 $40	_{-27}	^{+31}$	&$ 59_{-30}^{+20}$	\\
$\rm M_{tot}$  &  $0.2\pm0.2$& $0.5_{-0.4}^{0.3}$\\
\hline     
\hline     
  \end{tabular}\\
\label{tab:sb}
\end{table}

\section{Conclusion}
\label{sec:conclusions}
We use central velocity dispersion measurements for 17 members of the galaxy cluster A611 
as constraints to refine a SL model for the cluster. 
The inclusion of velocity dispersion measurements improves determination of the mass associated with galaxies in the cluster relative to the mass contained in the extended dark matter halo. Without the spectroscopically determined $\rm \sigma_{sp} $, the mass associated with the galaxies is overestimated by a factor $\sim1.5$, and consequently the mass of the large scale dark matter is underestimated by $\sim5\%$. \\

In contrast with the cluster A383 \citep{Monna2015} where the use of central velocity dispersions substantially improves constraints on the truncation radii of cluster galaxies, there is little change in these constraints in the case of A611.
Errors  in the truncation radii for galaxies in A611 are $\sim75\%$ comparable with the errors from the SL modelling without using $\rm \sigma_{sp}$. This result for A611 relative to A383 is due to both the size of the samples of cluster members with measured velocity  and the number of lensed features. In A383  \citep{Monna2015} there are  $\sim10$ galaxies with measured $\rm \sigma_{sp}$ inside the critical lines and the constraints on the truncation radii  improve by $\sim50\%$. In A611,  there are only 5 galaxies inside the region  probed robustly by SL. Furthermore, A611 has only 3 systems of multiple images in the cluster core whereas A383 has 10. \\

SB reconstruction of the tangential giant arc associated with A611 does provide additional
constraints on 5 cluster members projected near the arc. The galaxies G2 and G4 get velocity dispersions from the SB analysis which are consistent at the $1\sigma$ confidence level with the respective spectroscopic measurements. Overall the parameters of the galaxies are in agreement with the scaling relation derived in the pointlike analysis. 
In contrast two of these galaxies depart substantially from the galaxy scaling relation. Their small truncation radii may reflect  differing stripping history among individual cluster members \citep[][]{Warnick2008}.

More extensive samples of spectroscopically measured velocity dispersions for members of
a set of clusters will eventually provide a platform for refining the relative contribution
of the cluster members and the dark matter halo to the overall cluster mass distribution as a function of total cluster mass and as a function of the evolutionary state of the cluster. Combined with large redshift surveys of the central cluster region they promise insight into the stripping processes that govern the evolution of galaxies in dense cluster environments.

\section*{Acknowledgements}
We thank Andrew Newman and David Sand for generously providing their Hectospec spectra of A611 so that we could extract central velocity dispersions of additional cluster members.
We also thank  Aleksi Halkola, who provided the tool \texttt{GLEE} used to perform this analysis. Finally we thank Megan Donahue, Brenda Frye, Claudio Grillo and Massimo Meneghetti for their comments and contributions to this work. 
This work is supported by the Transregional Collaborative Research Centre TRR 33 - The
Dark Universe and the DFG cluster of excellence ``Origin and Structure of the Universe". 
The CLASH Multi-Cycle Treasury Program (GO-12065) is based on observations made with the NASA/ESA Hubble Space Telescope. The Space Telescope Science Institute is operated by the Association of Universities for Research in Astronomy, Inc. under NASA contract NAS 5-26555. The Dark Cosmology Centre is funded by the DNRF. Support for AZ is provided by NASA through Hubble Fellowship grant \#HST-HF-51334.01-A awarded by STScI. The Smithsonian Institution supports the research of DGF, MJG, and HSH.
\addcontentsline{toc}{chapter}{Bibliography}
\bibliographystyle{mn2efix}
\bibliography{a611}

\appendix
\section{MCMC Sampling of the pointlike models}
\label{sec:A1}

\begin{figure*}
\centering
\includegraphics[width=17cm]{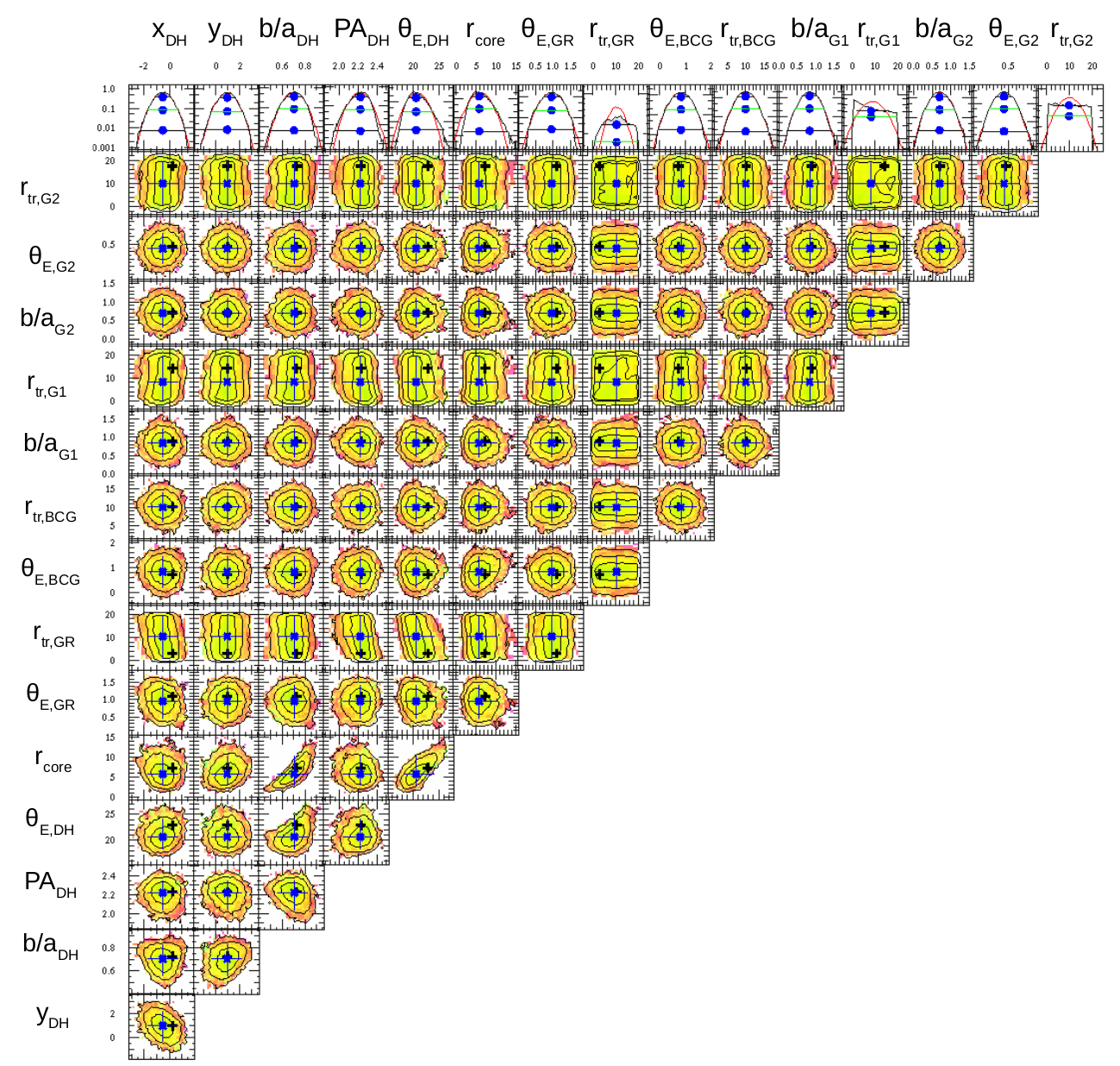}
\caption{\small MCMC sampling of the  parameters for the DH, BCG, GR and the galaxies G1 and G2 close to the giant arc. The coordinates x$_{\rm DH}$ and y$_{\rm DH}$ are in arcseconds with respect to the BCG position. Position angles are in radiants. The radii $r_{\rm core}$  and $r_{\rm tr}$ are in arcseconds. The Einstein radii ($\theta_E$) are also in arcseconds 
 for $D_{ds}/D_{d}=1$. }  
        \label{fig:mcmc_dh_g12}
\end{figure*}

\begin{figure*}
\centering
\includegraphics[width=16cm]{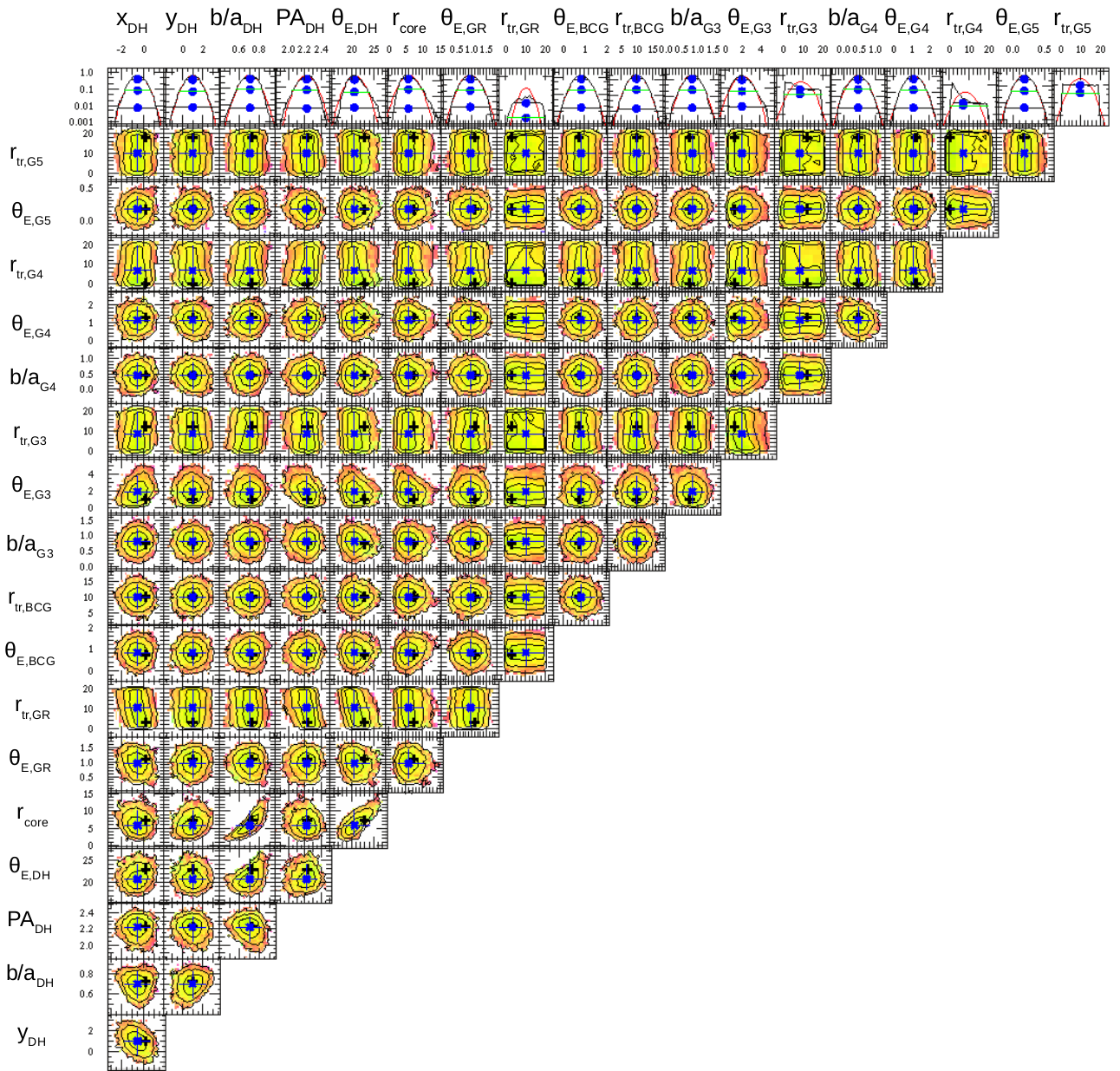}
\caption{\small Results of the MCMC analysis for the  parameters of the DH, GR and BCG mass components of A611, together with the mass parameters for the galaxies G3, G4 and G5. The units are as in the previous figure. }  
        \label{fig:mcmc_dh_g345}
\end{figure*}

In this section we present the Monte Carlo Markov Chain sampling of the parameters describing the mass components of A611 resulting from the pointlike analysis, presented in Sec.\,\ref{sec:A611}. The color scale correspond to 68.3\%  (green), 95.5\% (yellow) and 99.7\% (orange). The blue dot indicate the median of the distribution, the black cross is the best value. 
Fig.\,\ref{fig:mcmc_dh_g12} shows the MCMC sampling for the parameters of the smooth dark halo, the BCG, the reference galaxy GR and the galaxies G1 and G2. 
Fig.\,\ref{fig:mcmc_dh_g345} shows the sampling of the mass parameters for DH, BCG, GR together with the galaxies G3, G4 and G5. In both plots there is no degeneracy between the parameters of the galaxies in the neighborhood of the giant arc (G1-G5) and the mass components of the large scale halo, the BCG and the reference galaxy.

\bsp

\label{lastpage}

\end{document}